\documentstyle[11pt,aaspp4]{article}
\newcommand\beq{\begin{equation}}
\newcommand\eeq{\end{equation}}

\begin{document}

\title{$X$-ray Line Diagnostics of Hot Accretion Flows around Black Holes}
\author{Rosalba Perna\altaffilmark{1}, John Raymond and Ramesh Narayan}
\medskip
\affil{Harvard-Smithsonian Center for Astrophysics, 60 Garden Street,
Cambridge, MA 02138}

\altaffiltext{1}{Harvard Society of Fellows}

\begin{abstract}

We compute $X$-ray emission lines from thermal plasma in hot accretion
flows. We show that line profiles are strong probes of the gas
dynamics, and we present line-ratio diagnostics which are sensitive to
the distribution of mass with temperature in the flow.  We show how
these can be used to constrain the run of density with radius, and the
size of the hot region. We also present diagnostics which are
primarily sensitive to the importance of recombination versus
collisional ionization, and which could help discriminate ADAFs from
photoionization-dominated accretion disk coronae.  We apply our
results to the Galactic center source Sagittarius A$^*$ and to the
nucleus of M87. We find that the brightest predicted lines are within
the detection capability of current $X$-ray instruments.

\end{abstract}


\section{Introduction}

Many $X$-ray sources in the sky, such as stellar mass black holes in
$X$-ray binaries, and supermassive black holes in the nucleus of our
Galaxy and in other galaxies, are believed to be powered by gas
falling into the black hole. In the past few years,
advection-dominated accretion flows (ADAFs) have been successfully
used to model both spectra and luminosities in some of these systems
(Narayan \& Yi 1994, 1995; Abramowicz et al. 1995; see Narayan,
Mahadevan \& Quataert 1998 and Kato, Fukue, \& Mineshige 1998 for
reviews).  A feature of these models is that the gas is very hot,
approaching the virial temperature at each radius, and optically thin.
The emission from the inner region of the flow is dominated by
synchrotron radiation and Comptonization, and it has been studied in
detail by a number of authors (see Liang 1998 and Mushotzky, Done, \&
Pounds 1993 for reviews of the continuum spectral energy distribution
for these processes, and Narayan et al. 1998 for application to
ADAFs). On the other hand, the outer region of the flow, with
temperatures around $10^7-10^8$ K, is dominated by bremsstrahlung and
thermal $X$-ray line emission from hot, optically thin gas.

The electron temperature in an ADAF varies roughly as $T_e\sim 10^{12}
{\rm K}/r$, for $r\ga 10^3$ (Narayan \& Yi 1995b), where $r$ is the
radius in Schwarzschild units [2.95$\times 10^5 (M/M_\odot)$ cm, where
$M$ is the mass of the black hole].  Thus, the region of the flow
between $r\sim 10^4$ and $r\sim 10^5$ is the most important for
thermal line emission.  Narayan \& Raymond (1999) pointed out
significant diagnostic possibilities that the emission from this
region has.  In particular, they studied equivalent widths of emission
lines, and showed how these differ in various types of ADAF models.
Spectral fits to the emission from some accreting systems have
required the assumption that some mass is lost to winds during
accretion (e.g. Di Matteo et al. 1999).  The modification is described
by a parameter $p$ (defined in \S 2 below), which leads to a modified
density profile as a function of radius.  Another possibility is that
the ADAF may be convective, in which case the density profile is again
strongly modified (Narayan, Igumenshchev \& Abramovicz 2000, Quataert
\& Gruzinov 2000).  In effect, this density profile is similar to that
for an ADAF with a $p=1$ wind.  These possibilities lead to some
uncertainties in the details of ADAF models.  Another uncertainty is
related to the size of the ADAF region. In some systems (such as V404
Cyg, M87), the ADAF is believed to extend from the black hole horizon
to a transition radius $r_{\rm tr}$, beyond which the flow consists of
a thin disk plus a corona. But in other systems, e.g.  the Galactic
center source Sagittarius A$^*$ (Sgr A$^*$), there is no evidence for
an outer thin disk. Narayan \& Raymond did not study the effect of
varying $r_{\rm tr}$ on the lines.  Moreover, they assumed equilibrium
values for the ionic concentrations of the elements, and adopted
height-averaged equations for the ADAF. We improve on these aspects of
their model by computing time-dependent ionizations fractions, and
also adopting the general ADAF solution for the motion of the flow in
the $r\theta$ plane.

In this paper we present several diagnostics provided by thermal
emission from the hot plasma in the outer region of the flow.  Taken
together, these can set strong constraints on important properties of
the flow itself. We apply our results to Sgr A$^*$ and to the nucleus
of M87. 

In \S 3.1, we calculate line profiles. In ADAF models, these are
independent of the mass of the central black hole. However, they
depend on the temperature of the gas (or actually on how much gas
there is within each temperature range), on the amount of rotation of
the gas and on the inclination angle (i.e. the angle between the axis
of rotation of the flow and the direction of observation). We show
that there are several other diagnostics that constrain some of these
parameters independently (\S 3.2), such as line ratios and the slope
of the continuum.  Relative intensities of lines from low and high
ionization states are primarily sensitive to the relative density at
large and small radii.  In ADAF models, the run of gas density with
radius is uncertain, as is the position of the transition radius (if
there is one) to a thin disk plus a corona.  Some line ratios are very
sensitive to these aspects of the flow, and they can thus be
used to set constraints on them. This is particularly useful for the
case of SgrA$^*$, where both models with and without a wind are
consistent with the data (but require different assumptions for the
microscopic parameters of the flow; see, e.g., Quataert \& Narayan
1999), and there is no evidence of an outer disk. In this case, we
show that relative line intensities can be a useful, direct probe of
the amount of mass lost to a wind during the process of
accretion. Furthermore, we present other line-ratio diagnostics which
are primarily sensitive to the importance of recombination versus
collisional ionization, and we discuss how these can help discriminate
ADAFs versus photoionization-dominated accretion models.

In the case of M87, spectral data are much better fitted by an ADAF with a
wind (Di Matteo et al 1999). However, an uncertainty in the model for
this source is the location of the transition radius $r_{\rm tr}$
between the ADAF and an outer disk + corona.  We show that line ratios
are powerful diagnostics of this parameter.  Moreover, for this object
there is an observational estimate of the inclination angle of the
disk.  We then show that line profiles can be used to constrain the
amount of rotation of the flow and, hence, (see \S 2.1) learn about some
of its underlying microphysics.

\section{Model}

\subsection{Dynamics of the accretion flow}

We consider an axisymmetric ADAF model in spherical coordinates $r\;,\theta\;,
\phi$. In the equatorial plane ($\theta=\pi /2$) we assume that the velocity
components $v_r$, $v_\phi$ and the sound speed $c_s$ take the values given by
the analytic self-similar solution derived 
by Narayan \& Yi (1994).  For the variation
of the velocities with $\theta$, we make use of the work of Narayan \& Yi
(1995a), who have given analytic approximations for $v_r$, $v_\phi$ and
$c_s$, based on exact numerical calculations. In their solution $v_\theta =0$,
which we also assume. We thus have 
\beq
v_r(r,\theta) = -\frac{3\alpha}{5+2\epsilon} r\Omega_{\rm K}(r)\sin^2\theta\;,
\label{eq:vr}
\eeq
\beq
v_\theta=0\;,
\label{eq:vteta}
\eeq
\beq
v_\phi(r,\theta)=\left(\frac{2\epsilon}{5+2\epsilon}\right)^{1/2}
r\Omega_{\rm K}(r)\sin\theta\;,
\label{eq:vphi}
\eeq
\beq
c_s(r,\theta) =r\Omega_{\rm K}(r) [c_1-c_2\sin^2\theta]^{1/2}\;,
\;\;\;\; 
\label{eq:cs}
\eeq
where $c_1=2/5$, $c_2=(2/5)-2/(5+2\epsilon)$, and $\Omega_{\rm K}(r)=
(GM/r^3)^{1/2}$ is the keplerian angular velocity at radius $r$. We
express r in Schwarzschild units, $R_s=2.95\times 10^5
(M/M_\odot)$ cm, where $M$ is the mass of the black hole.  
Finally, the density is given by
\beq
\rho(r,\theta) = r^{-3/2}\rho_0[c_s^2(\theta)]^{-2.25}\;,
\label{eq:rho}
\eeq
as derived in Appendix A.  

The parameters $\alpha$ and $\epsilon$ contain information on the
microphysics of the flow: $\alpha$ is the viscosity parameter, for
which we assume a typical value for ADAFs of $\alpha=0.1$.  The
parameter $\epsilon$ is related to the ratio of specific heats of the
gas (Narayan \& Yi 1994).  At a macroscopic level, it describes the
amount of rotation of the flow.  In most of the calculations we adopt
$\epsilon=0.1$, but we also show how our results vary with $\epsilon$.

Given the radial scaling of $v_r$, the scaling of $\rho$ is uniquely
determined by the mass accretion rate $\dot{M}$
[Equation~(\ref{eq:6})].  In the simplest scenario, $\dot{M}$ is
constant throughout the flow. However, as explained in the
introduction, fitting the ADAF model to the spectra of some systems
requires that a fraction of the mass be lost to winds while
accreting. This implies an $\dot{M}$ which decreases inward.
Following the suggestion of Blandford \& Begelman (1999) and Di Matteo
et al. (1999), Quataert \& Narayan (1999) studied ADAF models with
winds; they assumed $\dot{M}\propto r^p$, and found that values of the
wind parameter $p$ between 0 and 0.5 are consistent with the spectral
data in several low-luminosity black holes. We consider this range of
$p$ in our models.

Another uncertainty of the models is the radial extent of the ADAF. As already
mentioned in \S 1, in some systems the ADAF region is
likely to extend up to a transition radius $r_{\rm tr}$, beyond which the flow
consists of a thin disk plus a corona. Following Esin, McClintock \& Narayan
(1997), in the coronal regions of these sources we set $\dot{M}\propto
r^{-1}$ (a somewhat arbitrary choice).  

To summarize, we adopt the following parameterization of
$\dot{M}$: 
\beq 
\dot{M} = \dot{m_0}\dot{M}_{\rm Edd}\left\{ \begin{array}{ll}
\left(\frac{r}{r_{\rm tr}}\right)^p & \hbox{if $\;\;r\le r_{\rm tr} $} \\
\left(\frac{r}{r_{\rm tr}}\right)^{-1} & \hbox{if $\;\;r\ge r_{\rm tr} $}
\\ \end{array}\right.\;,
\label{eq:mdot}
\eeq
where $\dot{M}_{\rm Edd}$ is the Eddington accretion rate. For a black
hole of mass $M$, we have $\dot{M}_{\rm Edd}=1.4\times 10^{18}(M/M_\odot)$ 
g s$^{-1}$.
The particular value of the constant $\dot{m_0}$ in any given source is
obtained by fitting the predicted flux for a given $\dot{M}$ to the 
observed $X$-ray spectrum. In Sgr A$^*$, we computed models with 
$p=(0,0.1,0.2,0.3,0.4,0.5)$, which gave
$\dot{m_0}=(0.000057,0.000077,0.00016,0.00017,0.00022,0.00026)$. 
In M87, for the same parameters, we found
$\dot{m_0}=(0.00086,0.0012,0.0018,0.0032,0.0061,0.01)$. 
It should be noted, however, that these estimates for the accretion
rate are still rather uncertain. In the case of Sgr A$^*$ the $X$-ray
emission from the accretion flow is not well constrained, both because
of the poor angular resolution of current observations and the large
uncertainty in the absorbing column to the source.  Some of the
$X$-ray flux might be due to diffuse gas surrounding the source.  For
M87, the uncertainty in the size of the ADAF region makes it unclear
whether the observed $X$-ray emission actually comes from the
accretion flow onto the central black hole (Reynolds et al. 1996), or
from extraneous point sources, or from the cooling flow which feeds
the accretion (Quataert \& Narayan 2000)\footnote{Observations with
the ACIS instrument on CXO should be able to eliminate these sources
of confusion. In fact, the resolution of this instrument is 1'' which,
for both Sgr A$^*$ and M87, corresponds to $r\sim 10^5$, that is the typical size
of the ADAF region.}. At any event, we also need to emphasize that,
while predictions for the {\em absolute} luminosity of the lines 
rely on the assumed value for $\dot{m_0}$, our results 
for the {\em relative} line luminosities and for line profiles are
unaffected by this uncertainty\footnote{Note that relative
line intensities are also  independent of the viscosity
parameter $\alpha$ and only very weakly dependent on $\epsilon$.}. 

A non-zero value of $p$ means that there is mass outflow.  In the
model developed by Quataert \& Narayan, this is accounted for by
taking an $\dot{M}$ which decreases with radius
[cfr. Eq. (\ref{eq:mdot})].  However, the motion of the flow is
assumed to remain the same as for $p=0$, namely 
an axially-symmetric rotation with purely
radial infall, as represented by Equations (\ref{eq:vr}) ---
(\ref{eq:vphi}).  The more general case, in which the outflow  also
affects  the dynamics, has been studied with numerical
simulations (Igumenshchev 1999; Igumenshchev \& Abramowicz 1999, 2000), 
but no analytical solution exists
yet.  In order to estimate the possible effects of an outflow on the
line profiles, we have developed an approximate model of the dynamics
of an ADAF with an outflow. This model leaves the rotational velocity
$v_\phi (r,\theta)$ the same as in the no-outflow model, 
but introduces a $\theta$ component for the velocity,
$v_\theta(r,\theta)$, while allowing $v_r(r,\theta)$ to be directed
both inwards and outwards. The details of this model are described in
Appendix B.

\subsection{$X$-ray spectra and line profiles}

Following Narayan \& Raymond (1999), we divide the flow into two regions
separated at a radius $r_{\rm in}=10^2$ in Schwarzschild units. 
In the inner part, the emission is
dominated by synchrotron, bremsstrahlung and Comptonization, and
these are all computed. In this region, line emission is neglected,
since the temperature is higher than $10^9$ K and therefore the
astrophysically abundant elements are fully ionized.  In the outer
part, on the other hand, synchrotron emission and Comptonization are
neglected, as these processes are steep functions of the
temperature. In this region thermal bremsstrahlung and lines dominate the
emission, and we compute them in detail. For the outer radius, we
 adopt $r_{\rm out}=10^5$ for all the models. This is consistent with the
evidence, in Sgr A$^*$, that the stars whose winds supply most of the
accreting mass are located at $r\ga$ few time $10^5$ (Coker \& Melia 1997;
Quataert, Narayan, \&  Reid 1999). 

We divide our flow into a grid in the $r\theta$ plane. At every point on
the grid we compute velocities and densities of the ADAF model as
described  in the previous section. The local temperature is determined
through the relation
\beq  
T(r,\theta)=\frac{\mu m_p}{k_{\rm B}}\;c_s(r,\theta)\;
\label{eq:temp}
\eeq where $\mu$ is the mean molecular weight of the gas (for which we
assume standard abundances), $m_p$ the proton mass, and $k_{\rm B}$
the Boltzmann constant. The $X$-ray spectrum in the outer region is
calculated using an extended version of the Raymond \& Smith (1977)
code. This computes bremsstrahlung, recombination, two-photon continua
and the emission in spectral lines.  Given the temperatures in the
flow, most of the line emission comes from H-like and He-like ions.
Included in the code are collisional excitation (see, e.g. Pradhan,
Norcross \& Hummer 1981 and Pradhan 1985), recombination to excited
levels of H-like and He-like ions (see, e.g. Mewe, Schrijver, \&
Sylwester 1980) and dielectronic recombination satellite lines (see,
e.g. Dubau et al. 1981 and Bely-Daubau et al. 1982).

Narayan \& Raymond showed that neither photoionization nor
photoabsorption is very important, and therefore we do not
consider them here. However, they also assumed ionization equilibrium,
which is somewhat marginal. We have removed this restriction in our
model. For each angle $\theta_i$ of the grid, the computation of the
ionization fractions of each element proceeds from the outermost to
the innermost shell at $r_{\rm in}$, with each parcel of gas being
followed while moving from a position $r_i$ to a position $r_{i-1}$
with velocity $v_r(r_i,\theta_i)$.  The initial condition for the
parcels of gas that start their inflow from the outermost radii is chosen to be 
that of ionization equilibrium at the local temperature. The subsequent
behavior of the gas is not particularly sensitive to this choice. 
If the gas were to start at a colder temperature, it would rapidly
approach the same state, due to the very fast ionization rates.

Figure 1 shows how significant the
departures from equilibrium are for the flows in SgrA$^*$ and M87. Here
we compare the equilibrium (dashed line) and non-equilibrium (solid
line) ionization fractions of the most important ions of iron.  We see
 that the departure from equilibrium is only marginal. The
overall tendency is a shift in the concentration peak of an ion inward
by log$\Delta r~ \approx 0.05$. A comparison between ionization times
and infall times shows that a large departure from equilibrium is indeed not
expected.  For example, the ionization rate of Fe XXV is
$C_i\approx 5\times 10^{-13}$ cm$^{3}$ sec$^{-1}$ (e.g. Arnaud \& Raymond 1992)
 at a temperature of $\sim 5\times 10^7$ K, where this
ion is most abundant.  In the case of Sgr A$^*$, for a density $n_e
\sim 5\times 10^3$ cm$^{-3}$ and velocity $v_r\sim 100$ km sec$^{-1}$
in the corresponding part of the flow
\footnote{This is around $r\sim$ few time $10^4$.  In the inner part
of the flow, the increase in the velocity --- which would cause a
larger departure from equilibrium --- is compensated by an increase in
both density and temperature, which acts in the opposite direction.},
we get an ionization time $t_{\rm ion}=1/n_e C_i\sim 4\times 10^8$
sec, smaller than the infall time $t_{\rm inf}\sim r/v_r\sim 7\times
10^9$ sec.  In the case of M87, the lower density makes $t_{\rm ion}$
about two orders of magnitude larger, but this is more than compensated by the
increase of about three orders of magnitude in the infall time (due to
the larger $M_{\rm BH}$).

Finally, we discuss the computation of line profiles.
Let $\vec{n}$ be the unit vector identifying the direction of an
observer with respect to the axis of rotation of the flow. The 
profile of a given line that the observer will measure is then given by
\beq
\Phi(v) = \int_0^{2\pi}d\phi\int_0^\pi d\theta\sin\theta\int_{r_{\rm in}}^{r_{\rm
out}}dr\,r^2 E_m(r,\theta)\;\Phi'(v;\, r,\theta,\phi)
\;\;\;\;{\rm ergs} \;{\rm s}^{-1}\; ({\rm km}\;{\rm s}^{-1})^{-1}\;,
\label{eq:prof}
\eeq
where $E_m(r,\theta)$ is the emissivity in the line under
consideration and
\beq
\Phi'(v;\, r,\theta,\phi) =\frac{1}{\sqrt{\pi}\Delta}
\exp\left\{-\frac{[v-v_{\rm los}(r,\theta,\phi)]^2}{\Delta^2}\right\}\;.
\label{eq:proff}
\eeq
Here $\Delta=\sqrt{2k_{\rm B}T/m_a}$ is the thermal width of an atom of mass
$m_a$, and $v_{\rm los}=\vec{v}\cdot\vec{n}$ is the component of the
velocity along the line of sight to the observer. If 
$\vec{n}=\sin i\,\hat{x}+\cos i\,\hat{z}$
($\hat{x}$ and $\hat{z}$ are
unit vectors in cartesian coordinates), then for the case
with no outflow (i.e. $v_\theta=0$), it is easy to see that
\beq 
v_{\rm los}(r,\theta,\phi)=v_r(r,\theta)\sin\theta\cos\phi\sin i
+ v_r(r,\theta)\cos\theta\cos i -
v_\phi(r,\theta)\sin i\sin\phi. 
\label{eq:vlos}
\eeq
The more general case with
$v_\theta\neq 0$ is treated in Appendix B.

\section{Results}

\subsection{Line profiles}

In Sgr A$^*$, there is no observational constraint on the inclination
angle $i$ between the line of sight and the axis of rotation of the
accreting gas. Figure 2 shows the profile of a strong line of iron for
different choices of $i$.  Note that the shape of the line is quite
sensitive to $i$.

For a proper interpretation of the dominant effects in determining
line broadening, we need to make some numerical estimates. First,
thermal broadening.  Given the range of temperatures in the flow
($\sim 10^7-10^9$ K), the resulting thermal widths are on the order of
100-150 km s$^{-1}$ for lines of ions mostly abundant in the outer
regions (such as O VIII, Si XIII), and on the order of 180-200 km s$^{-1}$
for ions produced in the inner region, such as Fe XXVI.  Thermal
widths must then be compared with the velocities due to the bulk motion of
the gas. Typical radial velocities have a maximum of $\sim 10^3$ km~s$^{-1}$ 
in the plane $\theta=0$,  and decrease to become zero at the poles,
whereas rotational velocities can have a maximum, in the plane of rotation, 
from a few $\times 10^3$ km s$^{-1}$ ($\epsilon\sim0.1$) up to  
$\sim 10^4$ km s$^{-1}$ ($\epsilon\sim 1$). 

When the inclination angle is very small, the rotational component has
no effect, and the only contribution from bulk motion comes from the
component $v_r\cos\theta$ (see Equ. (\ref{eq:vlos})), which gives an
average contribution roughly comparable to the thermal width. However,
for non-negligible values of $i$, it is the component $v_\phi\sin i$
that dominates the broadening of the lines (while the outer part of
the wings is still further broadened by a factor roughly proportional
to the thermal width $\Delta$)\footnote{This can be understood by
considering that a line profile is obtained by the convolution of many
gaussians, each centered around a value $v_{\rm los}(r,\theta,\phi)$,
and with a width $\Delta$.}.

Note that, if no independent information on the inclination angle is
available, the amount of rotation of the flow and the
inclination angle itself cannot be constrained separately. This
degeneracy can be broken in the case of M87, where observations
indicate an inclination angle of $i \sim 42^0$ (Ford et al. 1994).
Figure 3 shows the profiles of some of the strongest lines predicted
for this source.  Lines from elements in a high ionization state have
generally broader wings, as expected, because higher ionization states
are more populated at higher temperatures. The broadest line in the
figure is the $\lambda 1.78$ line of Fe XXVI.  Figure 4 shows, 
for M87, the effects of an increased rotation on the
profile of the line $\lambda 1.78$. In cases such as M87, where the
inclination angle is constrained by independent observations, line
profiles can thus provide powerful tests of the amount of rotation in
the flow, and, in turn, help us learn about some of its
underlying microphysics (see \S 2.1).

As already mentioned, a fit to the $X$-ray spectrum of M87 with an
ADAF requires the presence of a strong wind (Di Matteo et al. 1999),
which implies that some mass is lost to an outflow. Therefore we tried
to estimate the effect that such an outflow might have on the line
profiles. In Figure 5 we compare the profile of the line $\lambda
1.78$ with no outflow (solid line) with the profile of the same line
when there is an outflow (dotted and dashed lines).  The model we used for
the latter is described in Appendix B.  The two cases shown correspond
to different parameterizations of the density.  The dotted line is
computed with $a=0$ in Equation (\ref{eq:15}). This implies a constant
density (with respect to $\theta$) and a radial component of $v_r$
which is comparable for the infalling mass (largest in the plane
$\theta=\pi/2$) and the outflowing mass (largest at the poles
$\theta=0,\;\pi$). The dashed line is computed with the choice
$a=0.5$, correponding to lower density at the poles but higher
velocity for the outflow.

We find that the presence of an outflow does not affect line profiles
significantly.  This can be easily understood within the framework of
our model.  We assumed that the outflow mainly affects $v_r$ and
$v_\theta$, but not $v_\phi$, and that the $r$ dependence is the same
as in the standard model.  This resulted [cfr. Eq.(\ref{eq:14})] in a
$v_\theta$ rather smaller than $v_r$ and $v_\phi$ for typical ADAF and
wind parameters. The radial component of the velocity, on the other
hand, can be, contrary to the standard model of pure infall, both
positive and negative. However, the situation where the outflowing
mass has a velocity comparable to the infalling one is very similar to
the case of pure inflow. In fact, from the point of view of a distant
observer, since the gas is optically thin, inflow and outflow have
identical signatures if the velocities are identical. In the case in
which the outflow has a  higher velocity than the inflow, we do
expect a difference, but the outflow has a lower density, and
therefore its contribution to the total emission is not  very
significant
\footnote{This feature of our model --- that is the higher the velocity
in the outflow the lower its density --- has been imposed to mimic the
results of the numerical simulations.}. 

Needless to say, the validity of these results is limited to the
particular outflow model described in Appendix B.  If, for example,
the assumption that the outflow does not affect the rotation of the
gas does not hold true, then the outflow will affect the line
profiles, as Figure 4 shows.

Finally, note that optical depth effects, which could slightly modify the 
shape of line profiles, are not significant here. The optical
depth to a line is given by $\tau\sim 0.0106\,f\,N_{\rm ion}/\Delta\nu$,
where $f$ is the line oscillator strength and $\Delta\nu=\Delta v/\lambda$.
We computed $N_{\rm ion}$ numerically and found, for the two strongest lines
that we consider,
$\tau\sim 0.3$ for the OVIII $\lambda 18.21$ line, and $\tau\sim 0.02$ for the
Fe XXVI $\lambda 1.87 $ line, in the case of M87. For SgrA$^*$,
we found $\tau\sim 0.01 $ and $\tau\sim 10^{-3}$ for the same lines, respectively.

\subsection{Other diagnostics}

Figure 6 shows the predicted $X$-ray continuum luminosity of SgrA$^*$
for different values of the wind parameter $p$.  Note that the
observed flux from the region of the spectrum below 2 keV would be
highly absorbed (due to the high column density to this source), but
we show the luminosity in a wider range of energy for clarity.  The
figure shows that the higher the value of $p$, the less emission there
is at higher energies and the more at lower ones, thus making the
slope steeper. This effect was pointed out by Quataert \& Narayan
(1999).  The difference in emission is due to the fact that the
density is $\propto r^{-3/2+p}$, implying that, for higher $p$, there
is more gas at larger radii, where the temperature is lower.

As already mentioned in the introduction, ratios between lines from
low and high ionization states are very sensitive to the distribution
of density with temperature in the emitting plasma. In ADAF models,
this is essentially determined by the parameters $p$ and $r_{\rm tr}$
(see Eqs. (\ref{eq:mdot}) and (\ref{eq:7})).  Therefore, relative line
intensities can be used as diagnostics of these two parameters.  In
the case of the source Sgr A$^*$, there is no evidence of a transition
to an outer disk, hence line ratios provide a useful, direct probe of
the wind parameter $p$.  Having an independent way (i.e. unrelated to
spectral fits of the emission from the innermost part of the flow) to
constrain the wind parameter in ADAF models is an important issue, as
both models with and without winds fit the spectral data for Sgr
A$^*$, but require different assumptions for microscopic parameters of
the flow (Quataert \& Narayan 1999).

Figure 7 shows ratios between some strong lines of iron as a function
of the parameter $p$.  As $p$ increases, the flow becomes overall
colder (as more mass is taken away from the inner, hotter regions) and
the relative concentrations of the ions of iron consequently vary,
with lower ionization states becoming more populated. Ratios between
two lines from different ions mostly reflect variations in their
relative concentrations.  These ratios have the strongest dependence
on the distribution of mass with temperature (and therefore on $p$),
as it can be seen from the dashed line (showing [Fe XXV$ \lambda
1.855$]/ [Fe XXVI$ \lambda 1.780$]) and the long - dashed
line\footnote{Both ratios are rather insensitive to the importance of
recombination.}  (showing [Fe XXVI$ \lambda 1.780$]/[Fe XXV$
\lambda1.850$]) of Figure~7.  On the other hand, the ratio [Fe XXV$
\lambda 1.590$]/[Fe XXV$ \lambda1.850$] (dotted line) is sensitive to
the electron temperature, due to the difference in excitation
thresholds of the two lines.  The fact that it is pretty flat shows
that the bulk of the ions is always produced around the same
temperature, and there is not a significant departure from ionization
equilibrium.

Note that the ratio [FeXXV $\lambda 1.867$]/[FeXXV $\lambda 1.850$] is
very sensitive to the importance of recombination, and, in turn, to
the presence of a significant photoionizing flux. Recombination to the
excited levels results in a cascade which passes through the $\lambda
1.867$ transition. As a result, the intensity of such a line would be
strongly enhanced with respect to its value due to only collisional
excitation. When photoionization is important, the ratio [FeXXV
$\lambda 1.867$]/[FeXXV $\lambda 1.850$] is expected to be $\ga 1$.
Other elements have lines whose ratio would show the same behaviour,
such as the ratio [$\lambda 6.740$]/[$\lambda 6.648$] of the ion Si
XIII.  Photoionization in ADAF models was shown to be unimportant by
Narayan \& Raymond, but it would be significant in other cases, such
as the Compton-dominated corona models\footnote{Typical spectra of
photoionization-dominated emission are those from $X$-ray low-mass
binaries, and they show very different signatures from those discussed
here (see e.g.  the case of Cyg X3 discussed by Liedhal \& Paerels
1996). }. Therefore, the values of such ratios can be useful
diagnostics to discriminate ADAFs from these types of accretion
models\footnote{In photoionization-dominated accretion models,
however, the overall intensity of {\em all} lines would be lower, as
pointed out by Narayan \& Raymond (1999).}.

In the case of M87, applications of ADAF models give a much better fit
to the data for high values of the wind parameter ($p\sim 0.4-0.5$; Di
Matteo et al. 1999). The transition radius $r_{\rm tr}$ between the
inner ADAF and an outer disk + corona is, however, quite uncertain.
It may be possible to use relative line intensities to set constraints
on $r_{\rm tr}$.  Given the dependence of $\dot{M}$ on $r$ of Equation
(\ref{eq:mdot}), it can be seen that the smaller $r_{\rm tr}$, the
less gas there is at large radii, therefore resulting in an overall
hotter flow.  Figure 8 shows how temperature-dependent line ratios can
then be used to probe the size of the ADAF region and its transition
to a disk + corona.  The solid line shows the ratio [Fe XXVI $\lambda
1.780$]/[Si XIV$ \lambda 6.180$]; this is quite sensitive to the
relative emission from different regions of the flow: Fe XXVI is
produced in the very inner part whereas Si XIV is mostly abundant in
the middle and outer regions. A low value of $r_{\rm tr}$, with a
consequent high density in the inner part, produces an increase in the
abundance of Fe XXVI compared to Si XIV.  The ion O VIII, on the other
hand, is produced at lower temperatures compared to Si XIV, and this
is shown in the decrease of the ratio [Si XIV $ \lambda 6.180$]/[O
VIII$ \lambda 18.97$] ] (dotted line) with increasing $r_{\rm
tr}$. The other two ratios in Figure 8, that is [Si XIII $\lambda
6.648$]/[Si XIV $\lambda 6.180$] (dashed line) and [Fe XXV $\lambda
1.850$]/ [Fe XXVI $\lambda 1.780$] (dotted - dashed line), are mostly
sensitive to the relative ionic concentrations. As the flow is not too
far from equilibrium, these ratios are sensitive to the distribution
of mass with radius. If $r_{\rm tr}$ is small, more mass is
concentrated in the inner and hotter regions, thus making the higher
ionization states of the elements more populated. This point is
further illustrated by Figure 9, which shows the emissivity of some of
the strongest lines as a function of radius for two different values
of $r_{\rm tr}$, i.e. $r_{\rm tr}=10^4$ and $r_{\rm tr}=10^5$. While
the temperature profile is the same in both cases, the density profile
is different, with more mass in the outer region for the case of
$r_{\rm tr}=10^5$ (see Equ. (\ref{eq:mdot})) than for the case of
$r_{\rm tr}=10^4$. Lines produced in the outer region are therefore
enhanced compared to lines produced in the inner part of the flow, if
larger values of $r_{\rm tr}$ are assumed.

\subsection{Observational prospects}

Table 1 shows the luminosities of some of the most important lines for
both Sgr A$^*$ and M87. For each source, we considered one model
without wind ($p=0$), and one with a moderate wind ($p=0.4$).
Equivalent widths (EWs) for the same lines have been computed by
Narayan \& Raymond. For M87 and a wind model, the strongest lines at
both low and high energies have typical EWs $\ga$ a few tens of $\AA$.
This is the case also for the strongest iron lines in the models of
Sgr $A^*$.  In the case of this source, these would be the only
observable lines, due to the high column density ($N_{\rm H} \sim
6\times 10^{22}$ cm$^{-2}$), which causes a strong absorption of the
emission below about 2 keV. This is not a problem for M87, where the
column density is much lower [$N_{\rm H} = (2.1\pm 0.3)\times 10^{20}$
cm$^{-2}$, Sankrit, Sembach, \& Canizares 1999]. Given the distances
to these sources (8.5 kpc for Sgr A$^*$ and 16 Mpc for M87), the
brightest lines are already within the detection capability of the
ACIS detector on the {\em Chandra $X$-ray Observatory} (CXO).  With an
exposure of $10^5$ sec, the flux threshold is $\sim 2\times 10^{-14}$
erg cm$^{-2}$ sec$^{-1}$ around 6 keV (where most of the iron lines
are) and $\sim 6\times 10^{-16}$ erg cm$^{-2}$ sec$^{-1}$ at an energy
$\sim 0.6$ keV (where the strong O VIII line is). The higher-energy
threshold corresponds to a luminosity of $\sim 6\times 10^{38}$ ergs
s$^{-1}$ for M87 and $\sim 2\times 10^{32}$ ergs s$^{-1}$ for Sgr
A$^*$ (without correction for absorption), while the limiting flux at
0.6 keV corresponds to $L\sim 2\times 10^{37}$ ergs s$^{-1}$ for
M87. Lines in this range of the spectrum are highly absorbed for Sgr A$^*$.

For measurements of line profiles, the spectral resolution of HETG on
CXO is $\Delta E/E\sim 5\times 10^{-3}$ around 6 kev, while the LETG
spectral resolution is $\Delta E/E\sim 2.5\times 10^{-3}$ at about 0.6
keV, which corresponds to $\Delta v\sim 700$ km s$^{-1}$.  Given the
inclination angle inferred for M87, this resolution should be
sufficient to constrain the rotational velocity of the gas (see Figure
4), if a sufficient number of counts is available for the line under
consideration. We used the {\em Chandra Proposal Planning Toolkit} to
process the flux from the OVIII $\lambda 18.97$ line predicted for M87
(while taking also absorption into account) with the ACIS-S HETG/MEG1
detector, and found $\sim 4$ counts/ksec. This yields about 400 counts
after a typical integration time of 100 ksec. The statistics should
therefore be large enough to allow an estimate of the line widths,
though not a detailed profile measurement.  Excellent spectra for the
low energy lines will also be provided by {\em XMM}.

The above estimates are based on models which have been fitted to the
present best estimates of the $X$-ray fluxes of Sgr A$^*$ and M87.
Observations with {\em CXO} at high angular resolution will provide
more accurate $X$-ray fluxes, which may lead to modest revision in the
estimated line intensities.

\section{Conclusions}

We have shown that high-resolution observations of line profiles have
the potential to probe the dynamics of thermal gas in accretion flows
around black holes.  Line profiles are, essentially, indicators of the
temperature of the emitting plasma and of its velocity, but they also
depend on the angle between the line of sight to the observer and the
axis of rotation of the flow.  In the standard ADAF model, velocities
are independent of the mass of the central object and, consequently,
line profiles are also independent of it.  For non-negligible values
of the inclination angle, we find that line broadening is dominated by
the bulk rotational motion of the gas.  We have demonstrated that in
cases, such as M87, where there is independent evidence of the value
of the inclination angle, the amount of rotation of the flow can be
well constrained.

We have shown that further knowledge of properties of the flow can be
obtained through measurements of line ratios and the slope of the
continuum, which are sensitive to the distribution of mass with
temperature in the flow. In standard ADAF models, this is primarily
determined by the amount of mass lost to a wind, and by the size of
the hot region. We have presented our results for ADAFs applied to the
sources Sgr A$^*$ and M87. In the case of Sgr A$^*$, where there is no
evidence of an outer disk, we have illustrated how relative line
intensities can provide strong indications of the mass lost to a wind
during accretion.  This is particularly useful for this source, where
the wind parameter is still uncertain and cannot be constrained by
spectral fits.  In the case of M87, where models with winds provide a
better fit to the data, we have showed that line ratios can be
powerful probes of the transition radius between the ADAF and an outer
disk + corona.  Finally, we have presented line ratio diagnostics that
can help distinguish ADAFs from photoionization-dominated accretion
models. We find that the strongest lines are within the detection
capability of current $X$-ray instruments.

\newpage

\centerline{APPENDIX}

\appendix

\section{$\theta$ dependence of the density profile}

Under the assumption that $v_\theta$=0 and the viscosity terms can be
neglected, the equation of hydrostatic equilibrium in the $\theta$ direction
reduces to
\beq
\frac{1}{r}\frac{dP}{d\theta}=\rho\;v^2_{\phi}\frac{\cot\theta}{r}\;.
\label{eq:1}
\eeq
Recalling that $p=c_s^2\rho$, with
$\rho(r,\theta)=\rho(\theta)r^{-3/2}$, Equation (\ref{eq:1}) gives
\beq
\frac{1}{\rho(\theta)}\frac{d}{d\theta}[\rho(\theta)c_s^2(\theta)]
=v_\phi^2\cot\theta
\label{eq:2}
\eeq
Using Equations~(\ref{eq:vphi}) and (\ref{eq:cs}), this becomes 
\beq
\frac{d\rho}{\rho}=(c_3+2c_2)\frac{\sin\theta\cos\theta}
{c_1-c_2\sin^2\theta}d\theta\;,
\label{eq:3}
\eeq
having defined $c_3\equiv 2\epsilon/(5+2\epsilon)$.
Equation~(\ref{eq:3}) can be integrated to yield 
\beq
\rho(\theta)=\rho_0\left(\frac{1}{c_1-c_2\sin^2\theta}\right)^{
\frac{c_4}{2c_2}}\;
\label{eq:4}
\eeq
where $c_4\equiv c_3+2c_2=3.6\epsilon/(5+2\epsilon)$.
Finally, noticing that $c_4/2c_2=2.25$ and using Equation~(\ref{eq:cs}),
we get
\beq
\rho(\theta)=\rho_0(c_s^2)^{-2.25}\;.
\label{eq:5}
\eeq
The constant of integration $\rho_0(r)$ depends on the mass
accretion rate through the equation (e.g. Narayan \& Yi 1995a)
\beq
\dot{M}(r)=-\int 2\pi r^2\sin\theta\rho(r,\theta)v_r(r,\theta)d\theta\;
\label{eq:6}
\eeq
which yields
\beq
\rho_0(r)=-\frac{\dot{M}(r)}{2\pi f\sqrt{GM}}\;,
\label{eq:7}
\eeq
where $f\equiv [-3\alpha/(5+2\epsilon)]\int_0^\pi d\theta 
(c_s^2)^{-2.25}\sin^3\theta$.

\section{Simplified model of an ADAF with an outflow}

We consider a self-similar flow in which the $r$ dependences
of density, sound speed and components of the velocity (including
$v_\theta$) are the same as in the pure ADAF with no outflow.
All variables are assumed to be independent of $\phi$ as before,
and the amount of rotation (expressed through $v_\phi$) is also left
unaltered.   
What we need to find is therefore the $\theta$ dependence of
$v_r,\;v_\theta,\;\rho,\;c_s$. We parameterize two of them, and 
find the other two from the continuity equation and the momentum
equation in the $\theta$ direction.
We look for a solution in which there is pure inflow in the 
equatorial plane $\theta=\pi/2$, and pure outflow at the poles. Therefore 
we require $v_\theta(0)=v_\theta(\pi)=0$. 
We also want a $v_\theta$ which changes
sign in going from the upper to the lower part of the plane of rotation.
The numerical simulations of Igumenshchev and Abramovicz
(1999) also suggest that 
higher velocities are accompanied by lower densities and vice versa.
The conditions required above are satisfied by the simple function
\beq
\rho(r,\theta)v_\theta(r,\theta)=\rho_0\;v_\theta^0\;
r^{p-2}\sin(2\theta)\;.
\label{eq:11}
\eeq
For the $r$ component of the momentum we look for something of the type 
$\rho(r,\theta)v_r(r,\theta)=\rho_0\;v_r^0\;r^{p-2}f(\theta)$; the
continuity equation
\beq
\frac{1}{r^2}\frac{\partial}{\partial r}(r^2\rho v_r) +
\frac{1}{r\sin\theta}\frac{\partial}{\partial \theta}
(\sin\theta \rho v_\theta) = 0
\label{eq:12}
\eeq
constrains the functional form of $f(\theta)$ together with
the relation between $v_r^0$ and $v^0_\theta$. That is, 
\beq
\rho(r,\theta)v_r(r,\theta)=\rho_0\;v_r^0\;
r^{p-2}[\cos(2\theta)+\cos^2\theta]\;
\label{eq:13}
\eeq
and 
\beq
v^0_\theta=\frac{v_r^0\;p}{2}\;.
\label{eq:14}
\eeq
At any given $r$, the value of $v^0_r$ represents the 
(pure infall) velocity in the $\theta=\pi/2$ plane, and thus
we choose it as in Equation~(\ref{eq:vr}). 
The value of $v^0_\theta$ is then determined by Equation~(\ref{eq:14}).
The constant $\rho_0$ is determined so that the total mass of the
flow is the same as in the case with no outflow. 

Next we parameterize the density. Igumenshchev \& Abramowicz (1999)
have presented numerical simulations 
with an outflow, which  show that the density is maximum in the plane of
rotation and decreases towards the poles.
This can be represented by the function
\beq
\rho(\theta)=\rho_0\;(1-a\; \cos^2\theta)\;,
\label{eq:15}
\eeq
where $\rho_0$ is the value of the density in the plane and 
$a\equiv (1-\rho_{\rm min}/\rho_0)$ expresses the degree of
variation of the density with angle. 
Finally, the sound speed is determined from the momentum
equation in the $\theta$ direction. Neglecting the viscosity
terms, this gives 
\beq
\rho\left[v_r\frac{\partial v_\theta}{\partial r} \;+\;
\frac{v_\theta}{r}\frac{\partial v_\theta}{\partial \theta}\;+\;
\frac{v_r v_\theta}{r}\;-\;\frac{v_\phi^2\cot\theta}{r}\right]\;=\;
-\frac{1}{r}\frac{\partial P}{\partial \theta}\;.
\label{eq:16}
\eeq
If the density does not decrease too steeply towards the poles
(in the standard ADAF with no outflow this is the case for
small $\epsilon$), then the first three terms in the left-hand side
of Equation~(\ref{eq:16}) are negligible compared to the fourth 
term. This can be seen using the condition (\ref{eq:14}) together
with the coefficients of Equations~(\ref{eq:vr}) and  (\ref{eq:vphi}), 
with $\epsilon=0.1$, $\alpha=0.1$. The $\theta$ component of the
pressure is then given by
$p(\theta)=\rho_0 c_3[-0.5\cos^2\theta+0.25a\cos^4\theta]$
+ const. The constant of integration is determined by imposing that
the sound speed $c_s=\sqrt{p/\rho}$ has the same value as in 
Equation~(\ref{eq:cs}) in the plane $\theta=\pi/2$.
Once the sound speed is obtained, the temperature can then be determined as
usual from Equation~(\ref{eq:temp}).

Finally, when $v_\theta\neq 0$, Equation~(\ref{eq:vlos})
for the component of the velocity
along the line of sight (needed for the computation of line
profiles) is generalized to
\begin{eqnarray}
v_{\rm los}(r,\theta,\phi)&=&v_r(r,\theta)\sin\theta\cos\phi\sin i
+ v_r(r,\theta)\cos\theta\cos i -
v_\phi(r,\theta)\sin i\sin\phi  \nonumber \\
&+& v_\theta(r,\theta)\cos\theta\cos\phi\sin i - 
v_\theta(r,\theta)\sin\theta\cos i\;.
\label{eq:los2}
\end{eqnarray}

\begin{figure}[t]
\centerline{\epsfysize=5.7in\epsffile{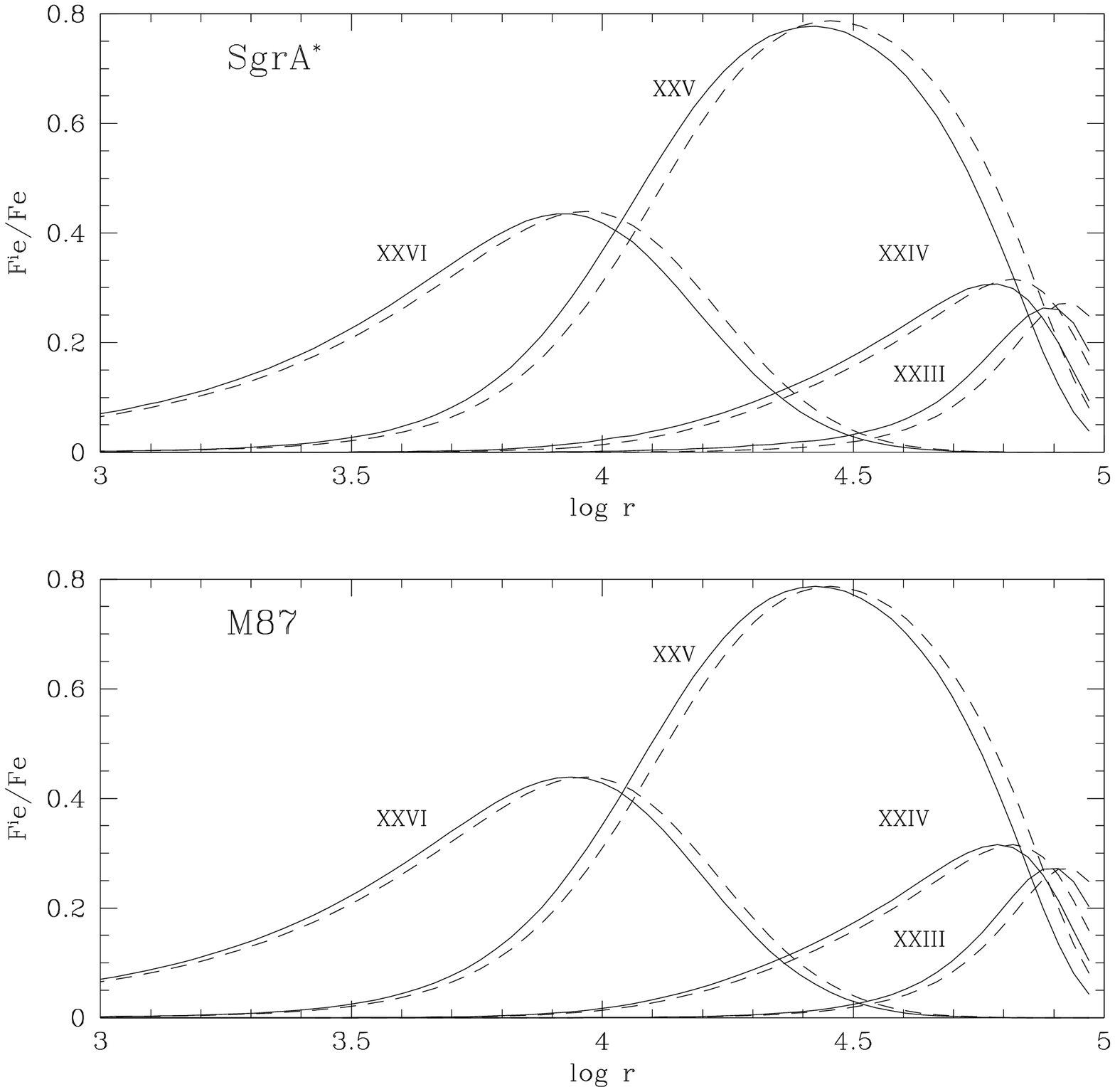}}
\caption{The effects of the departure from ionization 
equilibrium are shown for some
of the important ions of iron in the two sources Sgr A$^*$ and M87.  
In both panels, the dashed lines represent
the ionic concentrations at equilibrium, whereas the solid lines show
the non-equilibrium concentrations.}
\label{fig:1}
\end{figure}

\begin{figure}[t]
\centerline{\epsfysize=5.7in\epsffile{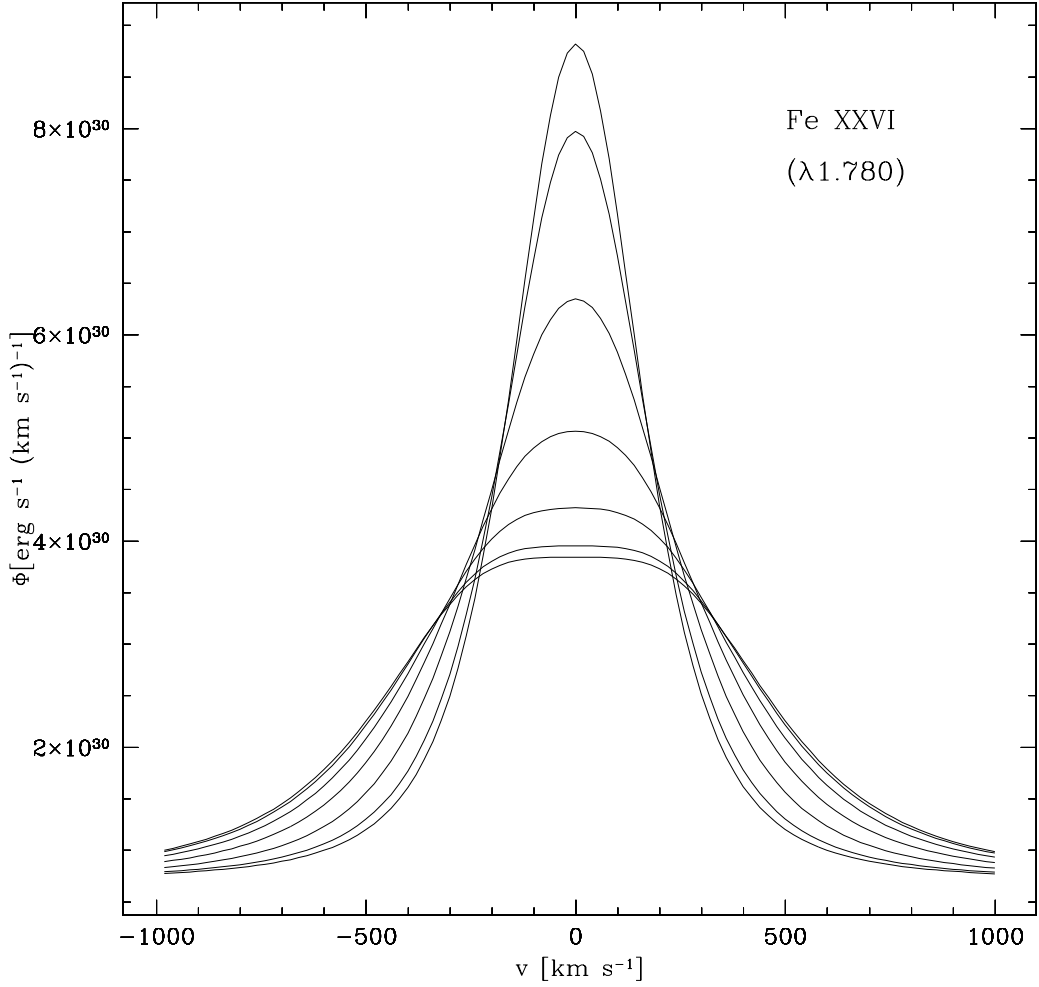}}
\caption{Line profiles for different choices of the inclination angle $i$ of
the line of sight with respect to the axis of rotation of the flow, for
the Galactic center source SgrA$^*$. The narrowest line corresponds to $i=0^0$;
the other lines (in order of increasing broadening) are computed 
with increments of $15^0$. The broadest line corresponds to $i=90^0$.
All the profiles have been computed for $p=0$ and $\epsilon=0.1$.}
\label{fig:2}
\end{figure}

\begin{figure}[t]
\centerline{\epsfysize=5.7in\epsffile{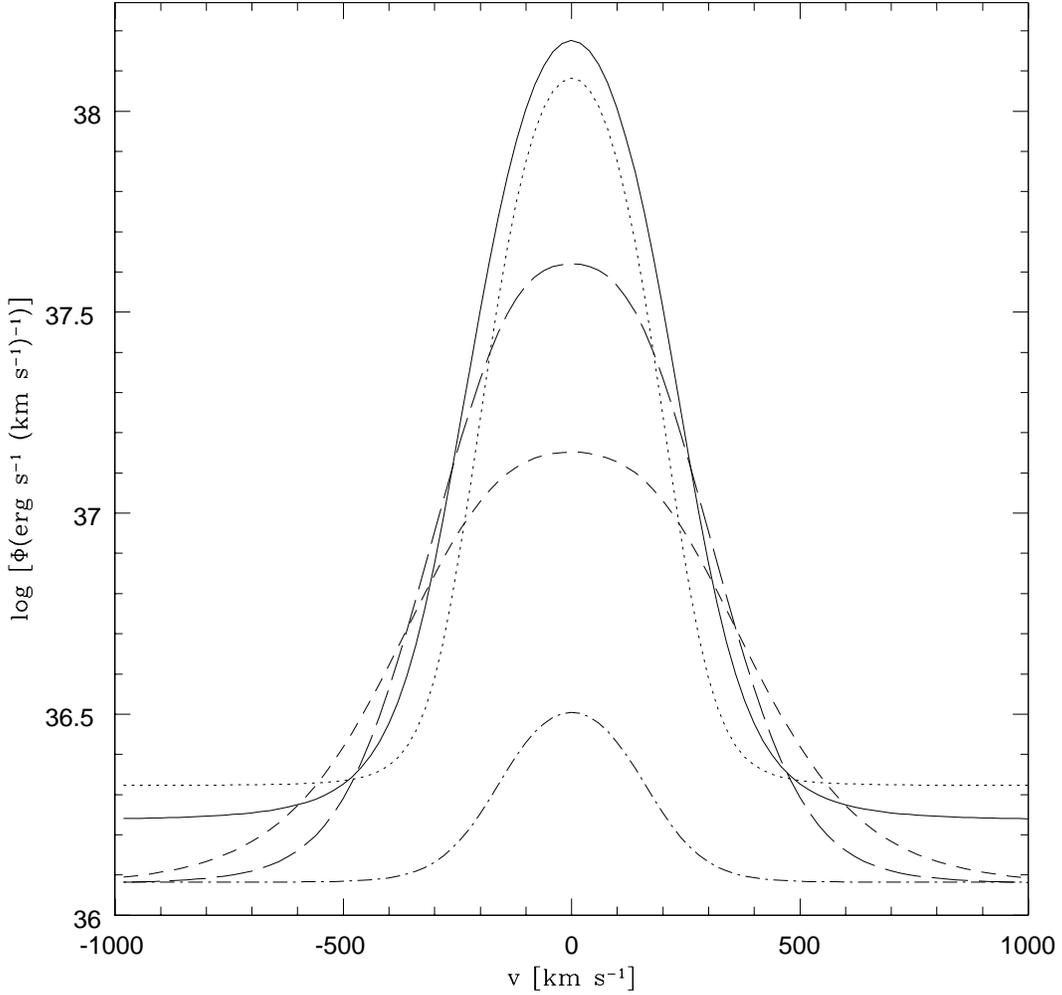}}
\caption{Line profiles of some of the strongest lines for the source
M87: O VIII $\lambda 18.97$ (solid line), Si
XIII $\lambda 6.648$ (dotted line), 
Fe XXIV $\lambda 1.861$ (dotted - dashed line), Fe XXV $\lambda 1.850$
(long - dashed line), Fe XXVI $\lambda 1.780$ 
(dashed line). The broadening of a line reflects the temperature of the
region in which it is produced; the intensity is proportional to the amount 
of gas at that temperature.    
Here the parameters are $p=0.5$, $\epsilon=0.1$, $r_{\rm tr}=10^5$
 and $i=42^0$.}
\label{fig:3}
\end{figure}

\begin{figure}[t]
\centerline{\epsfysize=5.7in\epsffile{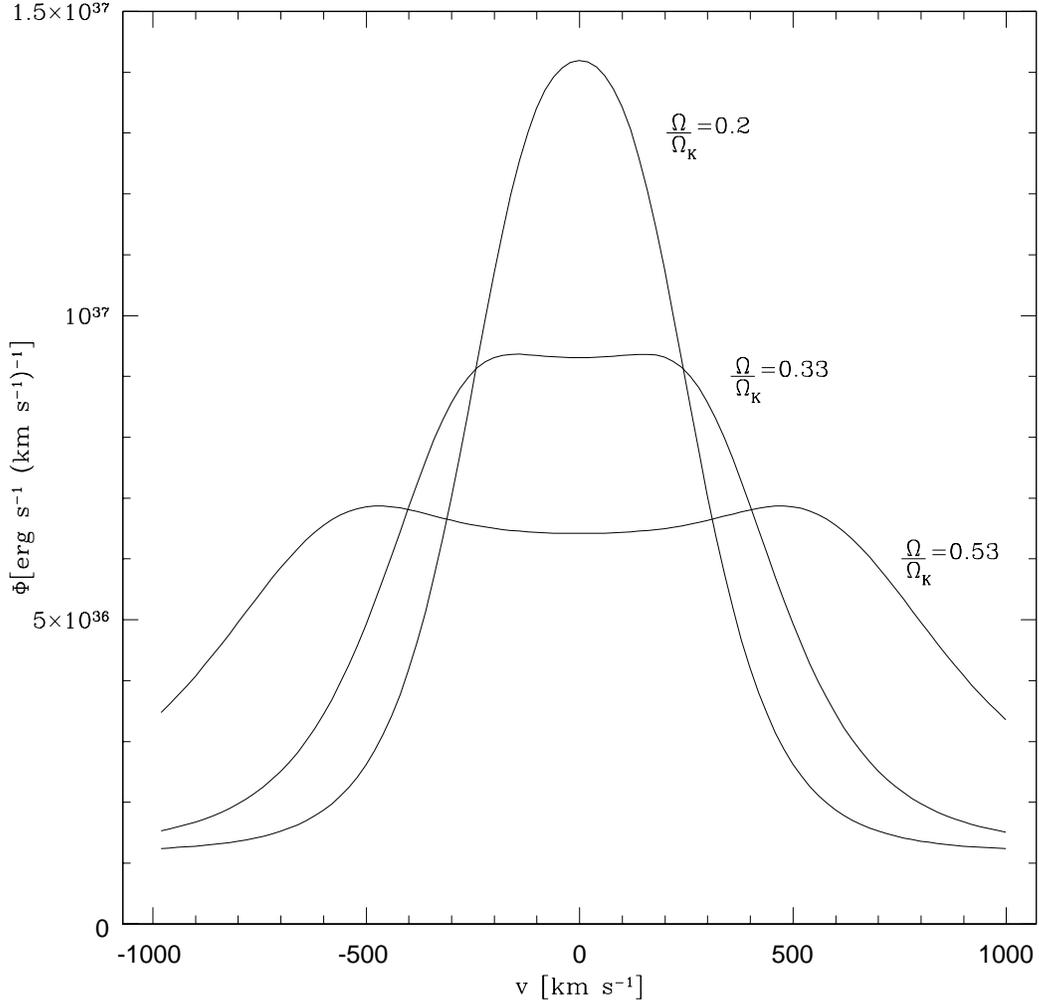}}
\caption
{Line profile of the line Fe XXVI $\lambda 1.780$ for the source M87.
Here we assume $p=0.5$, $r_{\rm tr}=10^5$, $i=42^0$, and show the
variation of the profile with the amount of rotation of the flow.
In the self-similar ADAF model that we are using, $\Omega/\Omega_{\rm K}=0.2$
corresponds to $\epsilon=0.1$, $\Omega/\Omega_{\rm K}=0.33$
to $\epsilon=0.3$, and $\Omega/\Omega_{\rm K}=0.53$ to $\epsilon=1$. }
\label{fig:4}
\end{figure}

\begin{figure}[t]
\centerline{\epsfysize=5.7in\epsffile{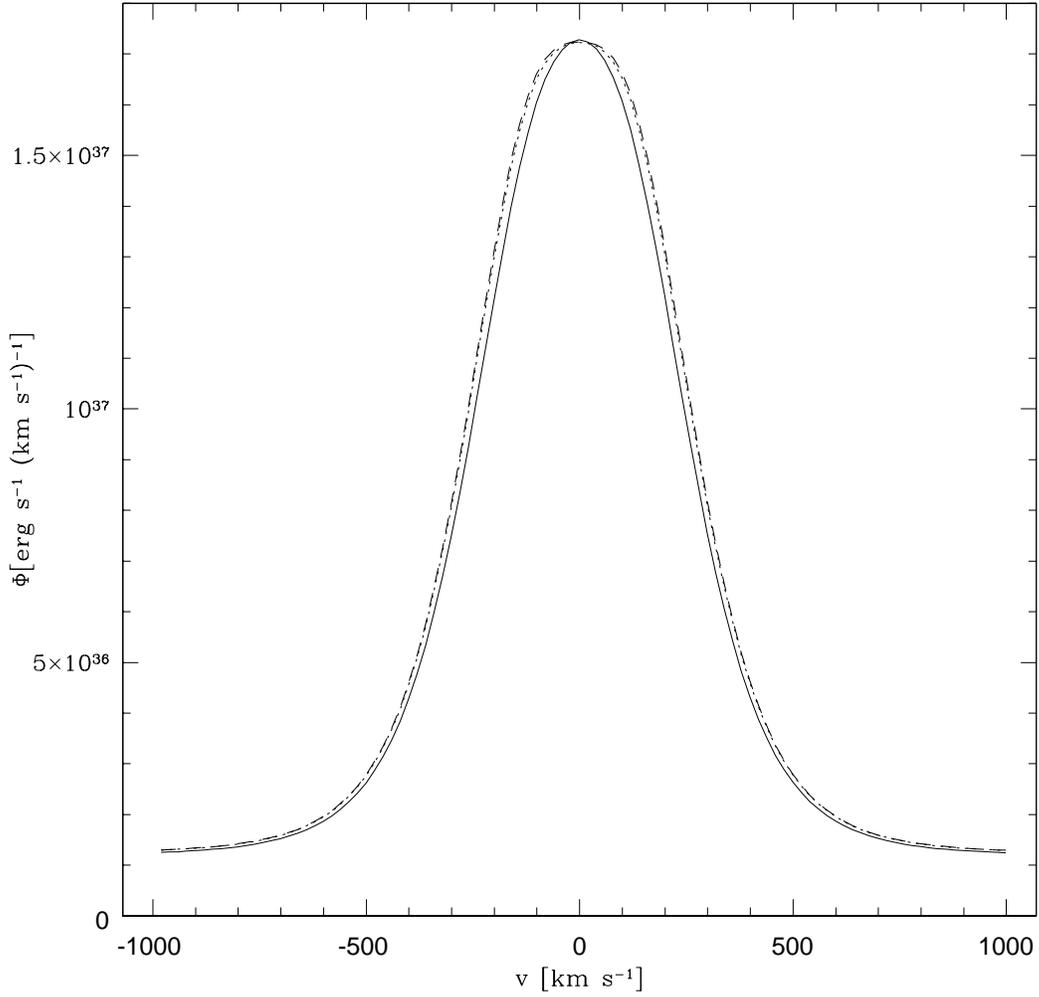}}
\caption{Effects of an outflow on the profile of the line Fe XXVI
$\lambda 1.780$.  The solid line is calculated with no outflow, while
the other two lines correspond to an outflow model with $a=0$ (dotted
line) and $a=0.5$ (dashed line).  The source here is M87 and the model
parameters are as in Figure 4 with $\epsilon=0.1$. In all cases, the
concentrations are calculated at equilibrium.  }
\label{fig:5}
\end{figure}

\begin{figure}[t]
\centerline{\epsfysize=5.7in\epsffile{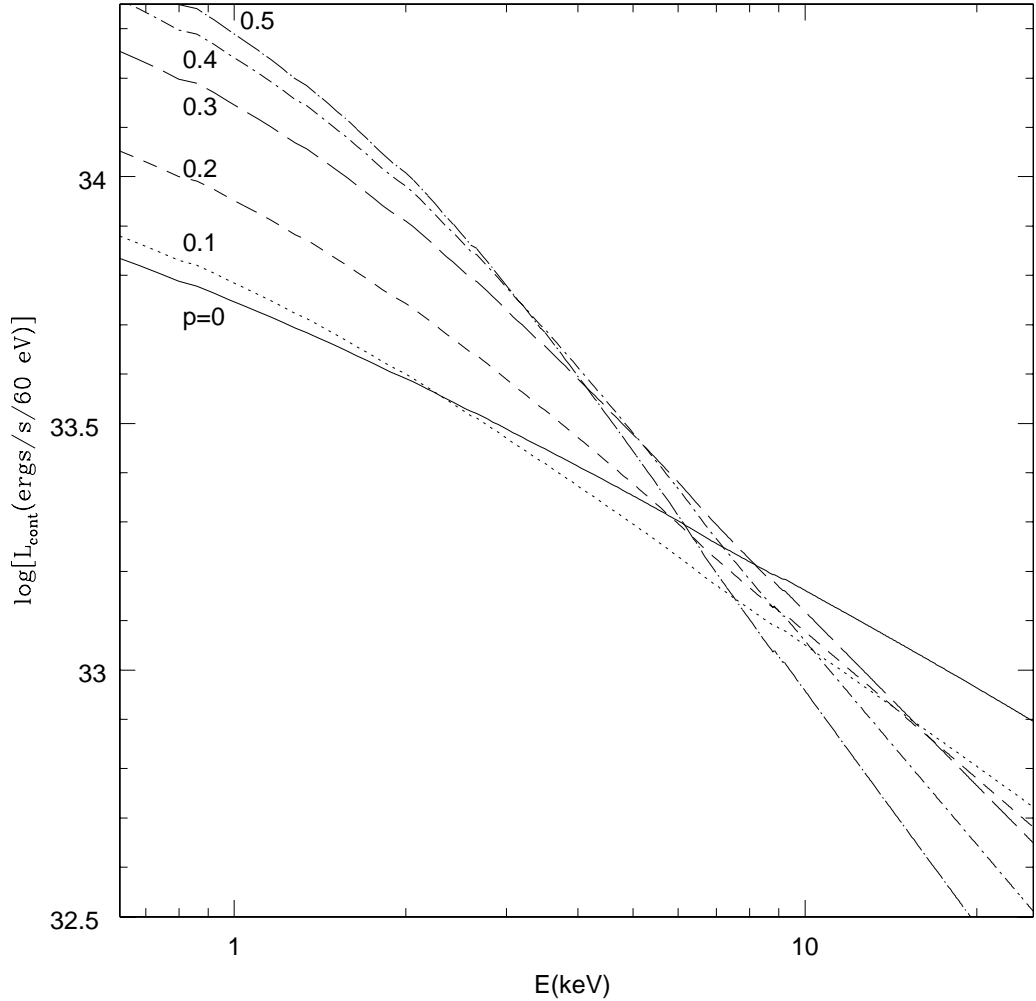}}
\caption{$X$-ray continuum of SgrA$^*$ for different values of the wind
parameter $p$. The higher the wind, the colder the gas, and consequently
the higher is the emission at low energies compared to high energies,
resulting in a steeper slope.}
\label{fig:6}
\end{figure}

\begin{figure}[t]
\centerline{\epsfysize=5.7in\epsffile{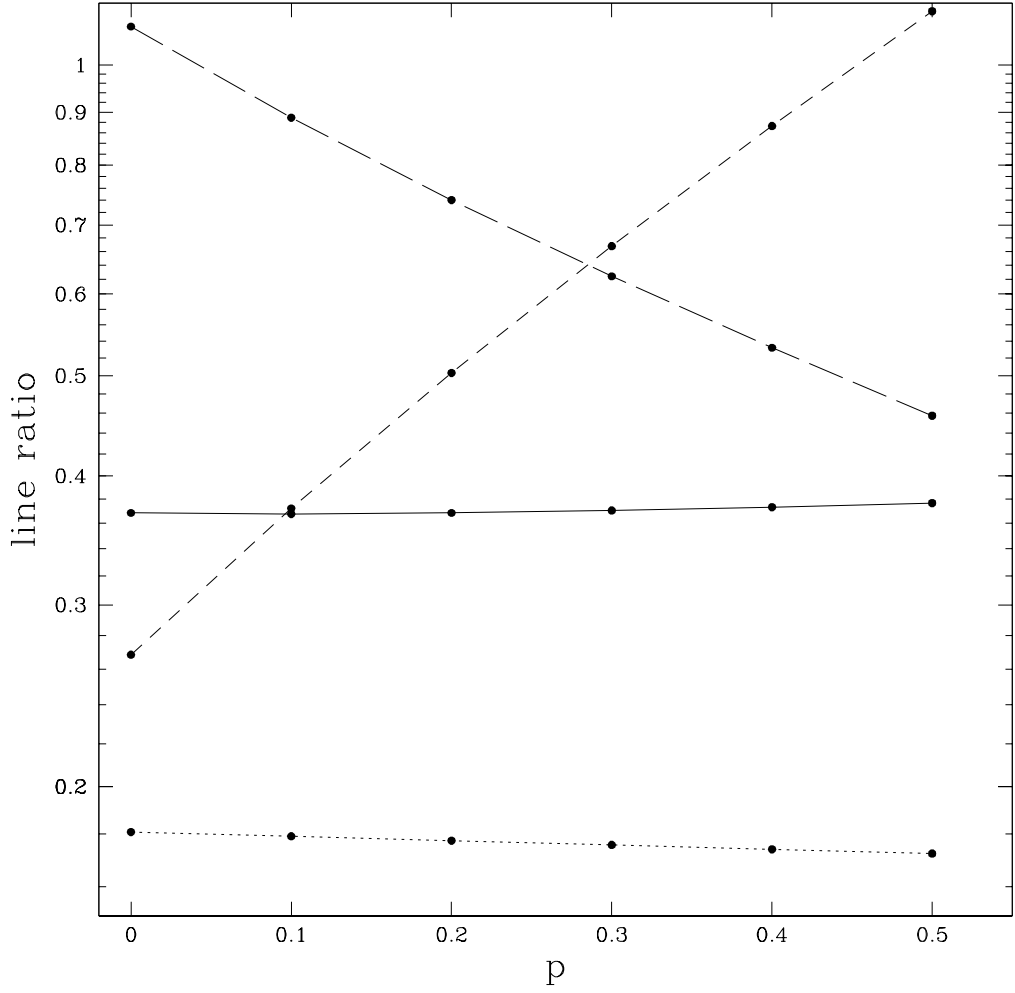}}
\caption{Line diagnostics of the wind parameter $p$ for SgrA$^*$.
The lines are the following: [Fe XXV $\lambda 1.867$]/[Fe XXV $\lambda
1.850$] (solid line), [Fe XXV $\lambda 1.590$]/[Fe 
XXV$ \lambda 1.850$] (dotted line),
[Fe XXV$ \lambda 1.855$]/[Fe XXVI$ \lambda 1.780$] (dashed line), 
[Fe XXVI$ \lambda 1.780$]/[Fe XXV$ \lambda1.850$] (long - dashed line).}
\label{fig:7}
\end{figure}

\begin{figure}[t]
\centerline{\epsfysize=5.7in\epsffile{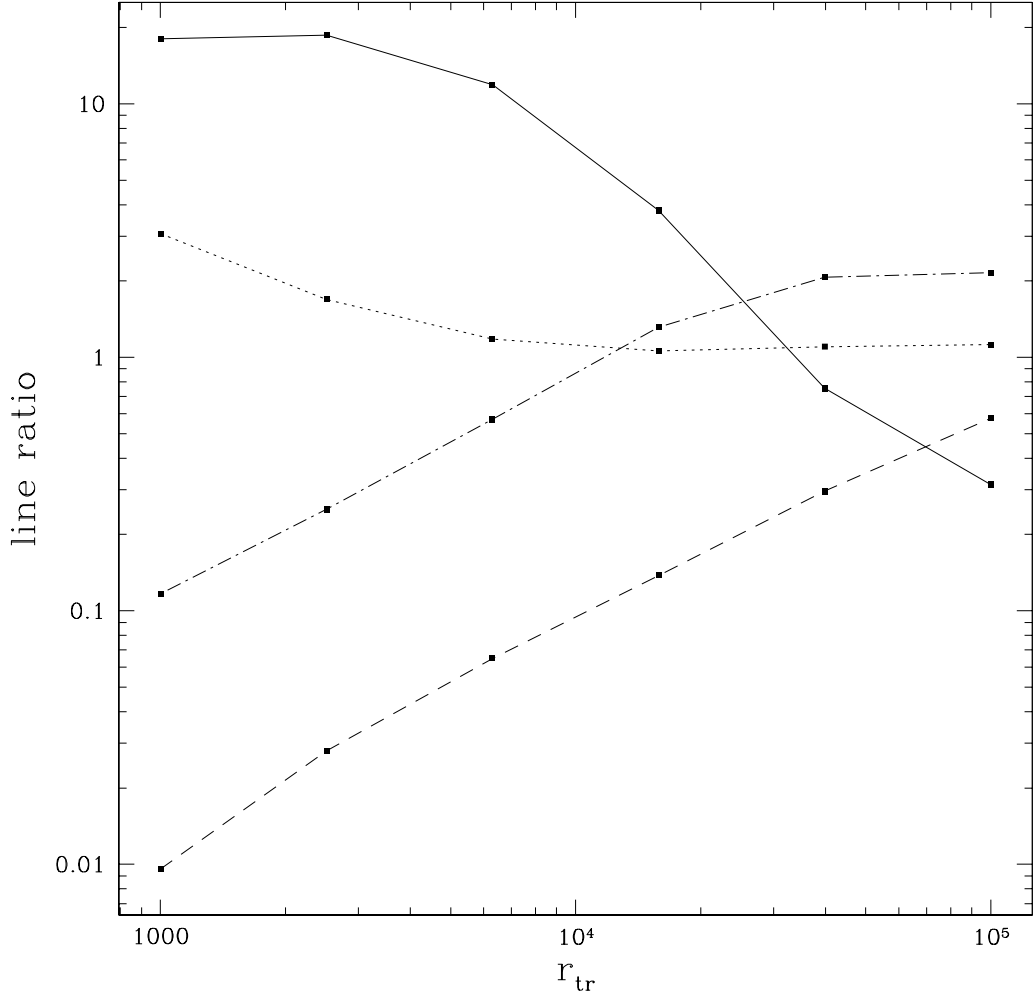}}
\caption{Line diagnostics of the transition radius between an ADAF and
a disk + corona for M87. The lines are the following: [Fe XXVI
$\lambda 1.780$]/[Si XIV$ \lambda 6.180$] (solid line), [Si XIV $
\lambda 6.180$]/[O VIII$ \lambda 18.97$] ] (dotted line), [Si XIII
$\lambda 6.648$]/[Si XIV $\lambda 6.180$] (dashed line), [Fe XXV
$\lambda 1.850$]/[Fe XXVI $\lambda 1.780$] (dotted - dashed
line). Here the wind parameter is $p=0.5$. }
\label{fig:8}
\end{figure}

\begin{figure}[t]
\centerline{\epsfysize=5.7in\epsffile{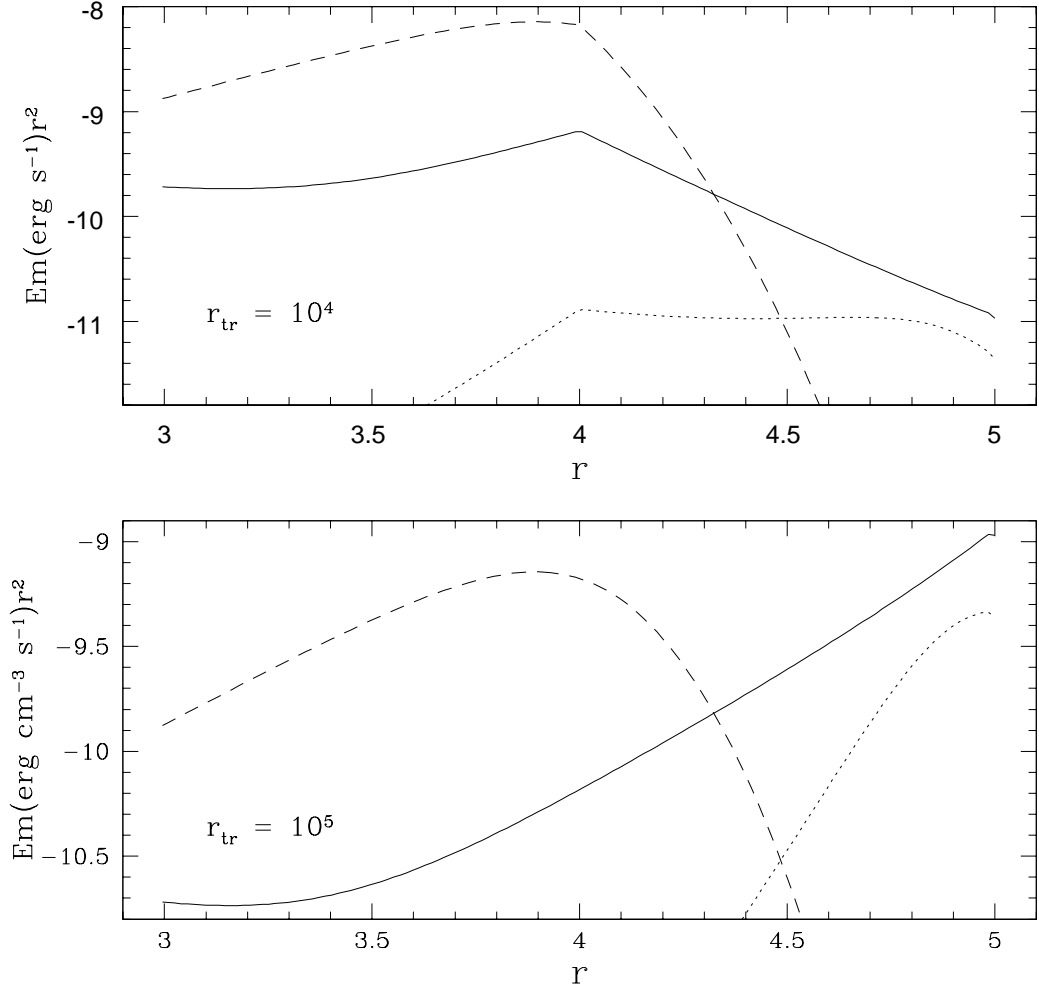}}
\caption{Emissivity in some of the strongest lines predicted for an
ADAF in M87: Fe XXVI$ \lambda 1.780$ (dashed line), O VIII$ \lambda
18.97$ (solid line), and Si XIII $\lambda 6.648$ (dotted line).  The
two panels show the effect of varying $r_{\rm tr}$. The temperature
profile is the same in both cases, but the density in the outer region
is higher for larger $r_{\rm tr}$.  This results in an enhancement 
of the total emission from the OVIII and Si XIII lines with respect to 
that from the Fe XXVI line.}
\label{fig:9}
\end{figure}

\newpage

\begin{tabular}{|c|c|c|c|c|} \hline
Line & SNW & SW & MNW & MW \\ \hline
O VII $\lambda21.60$     & $2.2\times10^{31}$& $2.6\times10^{32}$&$5.9\times10^{36}$ & $2.4\times10^{38}$\\
O VIII $\lambda 18.87$   & $2.3\times10^{33}$& $1.9\times10^{34}$&$6.2\times10^{38}$ & $1.7\times10^{40}$\\ 
Mg XI $\lambda 9.168$    & $1.3\times10^{32}$& $1.5\times10^{33}$&$3.4\times10^{37}$ & $1.3\times10^{39}$\\ 
Mg XII $\lambda 8.425$   & $8.1\times10^{32}$& $7.1\times10^{33}$&$2.1\times10^{38}$ & $6.4\times10^{39}$\\ 
Si XIII $\lambda 6.648$  & $6.7\times10^{32}$& $7.2\times10^{33}$&$1.7\times10^{38}$ & $6.6\times10^{39}$\\ 
Si XIV $\lambda 6.180$   & $1.7\times10^{33}$& $1.3\times10^{34}$&$4.6\times10^{38}$ & $1.2\times10^{40}$\\ 
S XV $\lambda 5.039$     & $6.1\times10^{32}$& $5.7\times10^{33}$&$1.6\times10^{38}$ & $5.1\times10^{39}$\\ 
S XVI $\lambda 4.727$    & $1.0\times10^{33}$& $6.1\times10^{33}$&$2.7\times10^{38}$ & $5.5\times10^{39}$\\ 
Ar XVII $\lambda 3.949$  & $4.1\times10^{32}$& $3.2\times10^{33}$&$1.0\times10^{38}$ & $2.9\times10^{39}$\\ 
Ar XVIII $\lambda 3.731$ & $5.8\times10^{32}$& $2.5\times10^{33}$&$1.5\times10^{38}$ & $2.3\times10^{39}$\\ 
Ca XIX $\lambda 3.173$   & $1.2\times10^{32}$& $7.8\times10^{32}$&$3.2\times10^{37}$ & $7.0\times10^{38}$\\ 
Ca XX $\lambda 3.020$    & $1.6\times10^{32}$& $5.3\times10^{32}$&$4.2\times10^{37}$ & $4.8\times10^{38}$\\ 
Fe XXIV $\lambda 1.861$  & $3.3\times10^{33}$& $1.1\times10^{34}$&$8.8\times10^{38}$ & $1.0\times10^{40}$\\ 
Fe XXV $\lambda 1.850$   & $1.2\times10^{33}$& $4.2\times10^{33}$&$3.3\times10^{38}$ & $3.8\times10^{39}$\\ 
Fe XXV $\lambda 1.867$   & $1.2\times10^{33}$& $4.5\times10^{33}$&$3.2\times10^{38}$ & $4.1\times10^{39}$\\ 
Fe XXV $\lambda 1.859$   & $6.1\times10^{32}$& $1.9\times10^{33}$&$1.5\times10^{38}$ & $1.7\times10^{39}$\\ 
Fe XXV $\lambda 1.590$   & $1.0\times10^{33}$& $5.3\times10^{33}$&$2.6\times10^{38}$ & $4.8\times10^{39}$\\ 
Fe XXV $\lambda 1.855$   & $5.1\times10^{31}$& $2.7\times10^{32}$&$1.3\times10^{37}$ & $2.4\times10^{38}$\\ 
Fe XXV $\lambda 1.780$   & $3.7\times10^{33}$& $6.0\times10^{33}$&$9.8\times10^{38}$ & $5.4\times10^{39}$\\ 
Fe XXV $\lambda 1.500$   & $7.4\times10^{32}$& $1.0\times10^{33}$&$1.9\times10^{38}$ & $9.2\times10^{38}$\\ 
Ni XXVII $\lambda 1.587$ & $1.8\times10^{32}$& $4.9\times10^{32}$&$4.7\times10^{37}$ & $4.4\times10^{38}$\\ 
Ni XXVIII $\lambda 1.530$& $1.8\times10^{32}$& $2.4\times10^{32}$&$4.9\times10^{37}$ & $2.2\times10^{38}$\\ \hline

\end{tabular}
\bigskip 

Table 1. Luminosities of some of the strongest lines for Sgr A$^*$ without
wind (SNW) and a moderate wind parameter of $p=0.4$ (SW), 
and for M87 without wind (MNW) and with $p=0.4$ (MW).  

\end{document}